\documentclass[conference,a4paper]{IEEEtran}
\usepackage{amsfonts}
\IEEEoverridecommandlockouts
\usepackage[a4paper, left=1.6cm, right=1.6cm, top=1.8cm, bottom=4.4cm, columnsep=0.21in]{geometry}

\ifCLASSINFOpdf
\else
\fi
\usepackage{multirow}
\usepackage{cite}
\usepackage{color}
\usepackage{graphicx}
\usepackage{amssymb }
\usepackage{algorithm}
\usepackage{algorithmic}
\usepackage{booktabs}
\usepackage{amsmath,bm}
\usepackage{colortbl}
\usepackage{makecell}
\usepackage{subcaption}
\usepackage[colorlinks]{hyperref}

\newtheorem{theorem}{Theorem}

\newtheorem{definition}{Definition}

\usepackage[percent]{overpic}
\def\BibTeX{{\rm B\kern-.05em{\sc i\kern-.025em b}\kern-.08em
    T\kern-.1667em\lower.7ex\hbox{E}\kern-.125emX}}
\begin{document}

\title{ A FAS Channel Fitting Strategy Using Extreme Value Distributions for Accurate Outage Performance Evaluation}

\author{
    \IEEEauthorblockN{Rui Xu\IEEEauthorrefmark{1}, Yinghui Ye\IEEEauthorrefmark{1}, Guangyue Lu\IEEEauthorrefmark{1}, Liqin Shi\IEEEauthorrefmark{1}, and Gan Zheng\IEEEauthorrefmark{2}}
    \IEEEauthorblockA{\IEEEauthorrefmark{1}Shaanxi Key Laboratory of Information Communication Network and Security,\\ Xi'an University of Posts and Telecommunications, Xi'an, China\\\IEEEauthorrefmark{2}School of Engineering, University of Warwick, Coventry, UK\\
    Emails: dhlscxr@126.com, connectyyh@126.com, tonylugy@163.com,\\ liqinshi@hotmail.com, gan.zheng@warwick.ac.uk}

    \vspace{-1.5em}

    \thanks{This work was supported in part by the National Natural Science Foundation of China under Grant 62571430 and 62471388, in part by the Natural Science Basic Research Program of Shaanxi under Grant 2026JC-QYCX-117, and in part by the Shaanxi Concept Demonstration Center under Grant 2025CG-GNYZ2-03.}
}\maketitle

\markboth{}
{Shi\MakeLowercase{\textit{et al.}}:}
\maketitle
\begin{abstract}
Modeling the channel in a single-antenna fluid antenna system (FAS) using extreme value distributions (EVDs) provides an accurate and tractable framework for FAS performance evaluation. When the objective of FAS channel fitting is outage probability (OP) evaluation, accurate characterization of the low-probability left-tail region becomes crucial, while existing fitting strategies that emphasize global fitting accuracy may fail to capture the critical tail behavior required for precise OP evaluation. In this paper, we propose an OP-oriented channel fitting strategy with a left-tail-sensitive target distribution and fitting criterion. Specifically, a combined EVD (CEVD) is introduced as the target distribution, where a generalized Pareto distribution (GPD) is employed to characterize the left tail and a generalized extreme value (GEV) distribution is used to model the global behavior. Furthermore, a modified mean-square-error (MMSE) criterion is developed, which employs logarithmic-domain errors to enhance sensitivity to left-tail discrepancies. Meanwhile, the evaluation points are constructed via uniform discretization on the logarithm of the cumulative distribution function, ensuring uniform sampling across all probability scales. This mitigates the under-representation of tail errors in the overall MMSE, which cannot be effectively addressed by error amplification alone due to the sparsity of tail samples. Simulation results demonstrate that the proposed fitting strategy significantly improves the OP evaluation accuracy in the ultra-low-OP regime.
\end{abstract}


\IEEEpeerreviewmaketitle

\section{Introduction}\label{sec:Intro}
\IEEEPARstart{F}{luid} antenna system (FAS) has emerged as a promising physical-layer technology for supporting ultra-reliable communications in sixth-generation networks \cite{10753482,11193779,11511867}. By densely deploying numerous antenna ports within a confined physical space, FAS exploits rich spatial diversity by dynamically switching the connection to the port with the most favorable channel, referred to as the FAS channel \cite{Xu2025PerformanceAO}. As a result, FAS can significantly enhance transmission reliability compared to conventional fixed antenna system.

A critical research topic in FAS is the accurate characterization of the achievable outage probability (OP), which is fundamentally determined by the statistical distribution of the FAS channel. Due to the dense port deployment, the channel envelopes across different antenna ports exhibit spatial correlation. This correlation leads to a multi-dimensional integral representation of the exact FAS channel distribution, making the derivation of a closed-form OP expression that is both accurate and tractable highly challenging \cite{Xu2025PerformanceAO, 10103838}. To address this challenge, extensive research efforts have been devoted to simplifying the spatial correlation structure \cite{9264694,wong2022closed,10623405,11192531}, reducing the complexity of multi-fold integrals \cite{10103838,10130117,11023237,10965723,10924151}, modeling the FAS channel using copula theory \cite{10253941,10678877,11230883}, and fitting the FAS channel distribution using extreme value distributions (EVDs) \cite{Xu2025PerformanceAO}. Among these approaches, the EVD-based fitting method proposed in \cite{Xu2025PerformanceAO} is particularly attractive due to its ability to provide relatively accurate OP expressions with low computational complexity.

In \cite{Xu2025PerformanceAO}, a maximum likelihood (ML)-based generalized extreme value (GEV) distribution fitting framework was proposed for accurate performance evaluation. Although the proposed framework accurately approximates the overall FAS channel distribution, non-negligible deviations can be observed in the low-OP region of practical interest (typically below $10^{-2}$), as shown in Fig. 8.
The limitation of the ML-based GEV fitting strategy for OP evaluation lies in its emphasis on global fitting accuracy of the FAS channel distribution, which may neglect the left-tail region that dominates low-OP events.
Specifically, although the GEV distribution accurately captures the asymptotic behavior of extreme values, it does not necessarily provide precise modeling of the left tail.
Besides, the ML criterion inherently prioritizes high-density regions of the distribution, leaving the low-probability left tail underrepresented.
Consequently, when channel fitting is intended for OP evaluation, both the target distribution and the fitting criterion should be designed with emphasis on left-tail fitting accuracy, thereby motivating the development of a left-tail-sensitive fitting strategy for accurate OP evaluation.

Toward this end, this paper proposes an OP-oriented FAS channel fitting strategy by jointly designing a left-tail-sensitive target distribution and fitting criterion. We first introduce a combined EVD (CEVD) as the target distribution, which integrates the generalized Pareto distribution (GPD) for accurate characterization of the sensitive left-tail region and the GEV distribution for global distribution modeling. Furthermore, a modified mean-square-error (MMSE) fitting criterion is developed for fitting the FAS channel distribution with the GPD. Specifically, the proposed criterion employs a logarithmic error metric to amplify discrepancies in the low-probability region, while the evaluation points are constructed through uniform discretization on the logarithm of the cumulative distribution function (CDF) to ensure uniform sampling across all probability scales. This mitigates the under-representation of tail errors in the overall MMSE, which cannot be effectively addressed by error amplification alone due to the sparsity of tail samples. Simulation results demonstrate that the proposed OP-oriented fitting strategy achieves accurate matching between the fitted distribution and the empirical FAS channel distribution, particularly in the left-tail region, thereby improving the OP evaluation accuracy.

\section{System Model and Channel Fitting Framework}\label{sec:Framework}
We consider a point-to-point FAS, in which a single fixed-antenna transmitter sends information to a receiver equipped with a fluid antenna (FA). The FA is connected to a single radio frequency chain and there are ${N}$ preset antenna ports evenly distributed along a straight line of length ${W}\lambda$, where $\lambda$ is the carrier wavelength and $W$ denotes the FA size.
Following prior works in \cite{Xu2025PerformanceAO,9264694,10678877}, the receiver activates the antenna port with the maximum channel envelope to receive signal, which can be expressed as $\left| {{h_{{\rm{FAS}}}}} \right| = \max \left\{ {\left| {{h_{\rm{1}}}} \right|, \left| {{h_{\rm{2}}}} \right|, \dots ,\left| {{h_N}}\right|}\right\}$. As such, the OP of the considered FAS is defined as ${P_{{\rm{out}}}} = \Pr \left\{ {\left| {{h_{{\rm{FAS}}}}} \right| \le \hat \gamma } \right\}$, where $\hat \gamma \! = \!\sqrt {{{{\gamma _{{\rm{th}}}}} \mathord{\left/ {\vphantom {{{\gamma _{{\rm{th}}}}} {\bar \gamma }}} \right. \kern-\nulldelimiterspace} {\bar \gamma }}}$, ${{\gamma _{{\rm{th}}}}}$ is the decoding signal-to-noise ratio (SNR) threshold, and $\bar \gamma \!= \!{P \mathord{\left/ {\vphantom {P {{N_0}}}} \right. \kern-\nulldelimiterspace} {{N_0}}}$ denotes the average transmit SNR with the transmit power $P$ and the noise power ${N_0}$.

  Clearly, the OP is determined by the distribution of  $\left|h_{\rm FAS}\right|$. However, as discussed in  \cite{9264694,wong2022closed,10623405}, the dense port deployment introduces spatial correlation across the
  channel vector ${\bf h} = \left[ h_1, \cdots, h_N \right]^T$, which follows ${\bf{h}} \sim CN\left({0,{\sigma ^2}{\bf{J}}} \right)$, where ${{\sigma ^2}}$ denotes the average channel power and ${\bf{J}}$ is the correlation matrix. The $\left( {i,j} \right)$-th entry of ${\bf{J}}$ is given by ${{\bf{J}}_{i,j}} ={J_0}\left( {2\pi W{{\left( {\left| {i - j} \right|}\right)} \mathord{\left/{\vphantom {{\left( {\left| {\bar n - \tilde n} \right|}\right)} {\left( {N - 1} \right)}}} \right.\kern-\nulldelimiterspace} {\left( {N - 1} \right)}}}\right)$, with ${J_0}\left(  \cdot  \right)$ denoting the zeroth-order Bessel function of the first kind. Such spatial correlation renders the exact statistical distribution of $\left|h_{\rm FAS}\right|$ mathematically intractable.

 To address the above issue, \cite{Xu2025PerformanceAO} develops an EVD-based fitting framework for the FAS channel, and the general fitting procedure is summarized as follows:

Step 1: Generate Monte-Carlo simulation samples of $\left|h_{\rm FAS}\right|$ according to $\left|h_{\rm FAS}\right| = \max \left\{ |h_1|, |h_2|, \dots , |h_N| \right\}$;

Step 2: Estimate the empirical distribution of $\left|h_{\rm FAS}\right|$ from the sample data;

Step 3: Fit the parameters of the target distribution using the selected fitting criterion based on the sample data or estimated empirical distribution.

Following this procedure, an approximate FAS channel distribution is obtained, based on which the OP can be evaluated. However, as mentioned in Section \ref{sec:Intro}, the ML-based GEV fitting strategy proposed in \cite{Xu2025PerformanceAO} may not capture the left-tail behaviors of the FAS channel, leading to inaccurate OP evaluation in the low-OP region.
To address this issue, this paper develops an OP-oriented FAS channel fitting strategy by jointly designing a left-tail-sensitive target distribution and fitting criterion to improve the OP evaluation accuracy in the low-OP region.

\section{OP-Oriented FAS Channel
Fitting Strategy}\label{sec:OP-driven}

\subsection{The CEVD}\label{sec:Combined}
In \cite{Xu2025PerformanceAO}, the FAS channel has been modeled by a GEV distribution, which characterizes the asymptotic behavior of the maxima/minima of a long finite sequence. While this approach provides an accurate global description, it fails to accurately capture the tail-oriented behavior that is critical for OP evaluation. To address this limitation, another class of EVDs, i.e., the GPD, which models the distribution of values exceeding a given threshold, offers a suitable alternative. Unlike the GEV distribution, which emphasizes the overall extreme value trend and is insensitive to sparse tail samples, the GPD treats threshold-exceeding values as extreme events, enabling a focused analysis of the tail behavior that governs ultra-low-probability events such as outage.
In what follows, we first introduce a brief background of the GPD.

\begin{theorem}\label{theorem1}
Let $X$ be a random variable with CDF ${F_X}\left( x \right)$. For a sufficiently low threshold $u$, define the lower-tail excess variable as $Y = u - X\left| {X < u} \right.$. Then, the conditional distribution of $Y$, i.e., $\Pr \left\{ {Y < y} \right\} = \Pr \left\{ {u - X < y\left| {X < u} \right.} \right\}$, can be approximated by the GPD with CDF given by\cite{9508780}
\vspace{-0.5em}
\begin{align}\label{1A}
{F_{{\rm{GPD}}}}\left( y \right) = 1 - {\left[ {1 + \frac{{{\xi _u}y}}{{{\sigma _u}}}} \right]^{ - {1 \mathord{\left/  {\vphantom {1 {{\xi _u}}}} \right. \kern-\nulldelimiterspace} {{\xi _u}}}}},
\end{align}
defined on $\left\{ {y:y > 0} \right\}$, where ${{\xi_u}}$ and ${{\sigma _u}}$ are shape and scale parameters of the GPD, respectively.\hfill {$\blacksquare $}
\end{theorem}

According to \emph{Theorem} \ref{theorem1}, for $x < u$, the left-tail distribution of $X$ below the threshold $u$ can be derived as
\begin{align}\label{2A}\notag
\Pr \left\{ {X < x} \right\} &= \Pr \left\{ {X < u} \right\} \cdot \Pr \left\{ {X < x\left| {X < u} \right.} \right\}\\\notag
&= \Pr \left\{ {X < u} \right\} \cdot \Pr \left\{ {u - X \ge u - x\left| {X < u} \right.} \right\}\\
&= {F_X}\left( u \right) \cdot \left[ {1 - {F_{\rm{GPD}}}\left( {u - x} \right)} \right].
\end{align}

We observe that the GEV and GPD distributions provide complementary tools for modeling the FAS channel envelope. Specifically, the random variable of interest is $X = \left| {{h_{{\rm{FAS}}}}} \right| = \mathop {\max }\limits_{1 \le i \le N} \left| {{h_i}} \right|$, which is a maximum-type random variable induced by port selection. Therefore, the GEV distribution is used as a baseline approximation to the overall distribution of $X$ in \cite{Xu2025PerformanceAO}. However, OP evaluation in the ultra-reliable regime is dominated by the lower tail of $X$, where the globally fitted GEV distribution may still exhibit noticeable deviations. To improve the lower-tail accuracy, the GPD is employed to model the lower-tail excess $Y = u - X\left| {X < u} \right.$, while retaining the GEV distribution as the baseline approximation for the remaining part of the FAS channel.
It should be emphasized that the proposed CEVD is not intended to be a strict asymptotic EVT law directly applied to the spatially correlated channel $\left\{ {\left| {{h_i}} \right|} \right\}_{i = 1}^N$. Instead, it is an EVT-inspired fitted distribution for the FAS channel $X = \left| {{h_{{\rm{FAS}}}}} \right|$. Although the port channels are spatially correlated, the correlation is accommodated through the fitted distribution parameters.

\begin{definition}\label{definition1}
Let $X = \left| {{h_{{\rm{FAS}}}}} \right| = \mathop {\max }\limits_{1 \le i \le N} \left| {{h_i}} \right|$ denote the FAS channel envelope.
For a predefined tail threshold $u$ corresponding to a target ultra-low probability, the CEVD is defined as
\vspace{-0.5em}
\begin{align}\label{3A}
{F_{{\rm{CEVD}}}}\!\left( x \right) \!= \!\left\{
{\begin{array}{*{20}{c}}
\!\!\!\!{{F_{\rm{GEV}}}\left( u \right)\! \cdot \! \left[ {1 \!- \!{F_{{\rm{GPD}}}}\left( {u - x} \right)} \right],x <
u}\\
\!\!\!\!{{F_{{\rm{GEV}}}}\left(x \right),\;\;\;\;\;\;\;\;\;\;\;\;\;\;\;\;\;\;\;\;\;\;\;\;\;\;\;\;\;\;x \ge u}
\end{array}}, \right.
\end{align}
where ${F_{{\rm{GEV}}}}\left( x \right) = \exp \left( { - {{\left( {1 + \xi \frac{{x - \mu }}{\sigma }} \right)}^{ - {1 \mathord{\left/ {\vphantom {1 \xi }} \right. \kern-\nulldelimiterspace} \xi }}}} \right),1 + \xi \frac{{x - \mu }}{\sigma } > 0$ is the GEV distribution, with $\xi$, $\mu$, and $\sigma>0$ representing its shape, location, and scale parameters, respectively.
The proposed CEVD is continuous at the connection point $u$, since $F_{\mathrm{GPD}}(0)=0$, which yields $\lim_{x \to u^-} F_{\mathrm{CEVD}}(x) =\lim_{x \to u^+} F_{\mathrm{CEVD}}(x)= F_{\mathrm{GEV}}(u)$.
\end{definition}

Note that the fitting accuracy of GPD is significantly influenced by $u$.
For simplicity, we choose $u$ as the value corresponding to a CDF value of ${10^{-2}}$, and thus $u$ can be obtained by solving ${F_{{\rm{GEV}}}}\left( u\right)={10^{ - 2}}$. This is because practical systems are primarily concerned with the OP below ${10^{-2}}$, and the GEV distribution provides accurate fitting above ${10^{-2}}$ observed in \cite{Xu2025PerformanceAO}.
The sensitivity of $u$ will be analyzed in Section \ref{sec:Results}.
In classical EVT, there are various methods determining the optimal threshold, such as the mean residual life method and the parameter stability method\cite{9508780}.
However, these classical methods are primarily developed to guarantee the asymptotic validity of the GPD approximation. In contrast, the objective of the proposed CEVD is not to strictly recover the asymptotic extreme-value limit, but rather to achieve accurate distribution fitting and OP evaluation for the FAS channel. Furthermore, classical methods determine $u$ via numerical algorithms for each system parameter setting, and may further require parameterizing $u$ for analytical tractability, which introduces extra modeling complexity.

\subsection{The MMSE Criterion}\label{sec:Revised}
In \cite{Xu2025PerformanceAO}, the ML criterion is employed for modeling the FAS channel distribution. However, the fitting emphasis of ML is implicitly determined by the sample distribution itself, where the abundant samples in high-density regions naturally dominates the likelihood function. As a result, it is difficult for the ML criterion to explicitly emphasize the outage-relevant low-probability region, leading to noticeable fitting deviations in the left-tail CDF. In contrast, the mean-square-error (MSE) criterion directly quantifies the discrepancy between the empirical and fitted CDFs, thereby providing the flexibility to emphasize fitting accuracy in specific probability regions through proper error weighting. In what follows, we first introduce the standard MSE criterion.

Consider a sample vector ${\bf{r}} = \left\{{{r_i}} \right\}_{i = 1}^I$ consisting of $I$ samples, and let the target distribution be fully characterized by a parameter vector ${\boldsymbol{\lambda }} = \left\{{{\lambda _j}} \right\}_{j = 1}^J$ with $j \in \left\{{1, \cdots ,J} \right\}$, where ${{\lambda _j}}$ denotes the $j$-th distribution parameter. Denote by ${{\hat F}_r}\left( r \right)$ and ${F_r}\left({r;{\boldsymbol{\lambda }}} \right)$ the empirical CDF and target model CDF, respectively. The standard MSE criterion is defined as
\vspace{-1.5em}
\begin{multline}\label{6A}
{\varepsilon _{{\rm{MSE}}}}\left( {{{\hat F}_r},{F_r};{\boldsymbol{\lambda }}} \right) \buildrel \Delta \over = \frac{1}{K}{\sum\limits_{k = 1}^K {\left( {{{\hat F}_r}\left( {{w_k}} \right) - {F_r}\left( {{w_k};{\boldsymbol{\lambda }}} \right)} \right)} ^2},
\end{multline}
where $K$ denotes the number of evaluation points used for CDF fitting, and ${{w_k}}$ denotes the corresponding sample values at which the fitting errors are evaluated.

When the objective of FAS channel fitting is the accurate OP evaluation, the standard MSE criterion may suffer from two limitations:

(1) In the left-tail region, the CDF values typically fall into an ultra-low-probability region, where the absolute differences between the fitted and empirical CDFs are usually very small. Although these discrepancies contribute little to the overall MSE, they may still result in significant deviations in the low-OP region. Therefore, it is necessary to enhance the sensitivity of the fitting criterion to left-tail discrepancies while relatively reducing the influence of less critical regions, so as to improve the fitting accuracy in the low-OP region.

(2) When the MSE criterion is constructed using samples drawn from the underlying channel distribution, only a limited number of samples fall into the left-tail region. As a result, even if the fitting errors in the left tail are amplified, their contribution to the overall MSE may still remain limited due to the sparse tail samples. Therefore, improving left-tail fitting accuracy requires not only enlarging the magnitude of tail discrepancies, but also increasing the effective participation of left-tail samples in the MSE evaluation.

To address these limitations, we design a tail-sensitive MMSE criterion that employs a logarithmic scale to amplify the left-tail errors and adopts a uniform sampling strategy in log-CDF domain to mitigate the ineffectiveness of error amplification caused by sparse samples.

\begin{definition}\label{definition2}
The MMSE criterion is defined as
\vspace{-1em}
\begin{multline}\label{7A}
{\varepsilon _{{\rm{MMSE}}}}\left( {{{\hat F}_r},{F_r};{\boldsymbol{\lambda }}} \right) \buildrel \Delta \over = \frac{1}{K}\sum\limits_{k = 1}^K {\left({\lg \left( {{{\hat F}_r}\left( {w'_k} \right)} \right)}\right.} \\
{\left. { - \lg \left( {{F_r}\left({{w'_k};{\boldsymbol{\lambda }}} \right)} \right)}\right)^2}.
\end{multline}
The evaluation point ${{w'_k}}$ is the sample value corresponding to a uniformly discretized log-CDF level, i.e., ${w'_k} = {{\hat F}_r}^{ - 1}\left( {{{10}^{{z_k}}}} \right)$, where ${z_k}$ denotes uniformly spaced points in the log-CDF domain $\left[ {\lg \left( {{{\hat F}_{{\rm{lower}}}}} \right), \lg \left( {{{\hat F}_{{\rm{upper}}}}} \right)}\right]$. Here, ${{{\hat F}_{{\rm{lower}}}}}$ and ${{{\hat F}_{{\rm{upper}}}}}$ denote the lower and upper bounds of the CDF, respectively.
\end{definition}

In Definition~\ref{definition2}, the logarithmic form of the errors serves to amplify left-tail discrepancies while reducing the relative influence of other regions, thereby addressing the insensitivity of conventional MSE metrics to tail errors. Furthermore, the evaluation points are constructed through uniform discretization on the log-CDF axis rather than directly using empirical samples. This construction ensures that each probability order of magnitude is represented by a comparable number of evaluation points, thereby significantly enhancing the contribution of the ultra-low-probability region in the MMSE evaluation and guiding the fitting procedure to prioritize the left-tail accuracy.

It should be noted that the errors can be weighted by ${1 \mathord{\left/ {\vphantom {1 {{F_r}\left( {r;{\boldsymbol{\lambda }}} \right)}}} \right. \kern-\nulldelimiterspace} {{F_r}\left( {r;{\boldsymbol{\lambda }}} \right)}}$ to emphasize the ultra-low-probability region. However, since ${{F_r}\left( {r;{\boldsymbol{\lambda }}} \right)}$ is extremely small in the tail, this weighting becomes excessively large, allowing individual rare samples to dominate the error and potentially destabilize the parameter estimates. In contrast, the proposed method does not directly amplify individual sample errors through probability-dependent weighting. Instead, the evaluation points are constructed via uniform discretization on the log-CDF axis, which ensures a balanced representation across different probability scales. As a result, the contribution of the ultra-low-probability region to the overall MMSE is significantly enhanced without allowing a few rare samples to dominate the fitting process.

\section{The FAS Channel Distribution Fitting and OP
Evaluation}\label{sec:OP evaluation}
In Section \ref{sec:OP-driven}, we take the CEVD as the target distribution, where the GEV distribution is employed to model the global region and the GPD is used to characterize the left-tail region. Since the ML-based GEV fitting has already been performed in \cite{Xu2025PerformanceAO}, its results can be directly utilized in this work. Therefore, this section focuses on the MMSE-based GPD fitting for the left tail below threshold $u$.

The procedure of the MMSE-based GPD fitting is summarized as follows:

\begin{itemize}
\item Generate Monte-Carlo simulation samples and estimate the empirical distribution of $\left|h_{\rm FAS}\right|$ following Steps 1 and 2 as described in Section \ref{sec:Framework}.
\item Determine the probability interval for GPD fitting $\left[ {{{\hat F}_{{\rm{lower}}}}, {{\hat F}_{{\rm{upper}}}}} \right]$;
    Uniformly discretize the logarithmic CDF levels to obtain ${z_k}$ over $\left[ {{{\hat F}_{{\rm{lower}}}}, {{\hat F}_{{\rm{upper}}}}} \right]$.
\item Map ${z_k}$ to the sample domain via ${w'_k} = {{\hat F}_r}^{ - 1}\left( {{{10}^{{z_k}}}} \right)$ to obtain the evaluation points;
\item Compute the MMSE over the set of evaluation points;
\item Determine the parameters of GPD by minimizing MMSE.
\end{itemize}
By following the above steps, we can obtain the GPD parameters (${{\xi _u}}$, ${{\sigma _u}}$) for a given $\left( {W,N} \right)$.
However, the fitting procedure must be repeated for different values of $W$ and $N$, leading to high computational complexity.

To improve the computational efficiency, the distribution parameters (${{\xi _u}}$, ${{\sigma _u}}$) are fitted as linear combinations of monomials over the prescribed probability interval $\left[ {{{10}^{ - 6}}, {{10}^{ - 2}}} \right]$ and the system parameter ranges $W \in \left[ {0.5,5} \right]$ and $\rho  \in \left[ {0.05,0.5} \right]$, where $\rho  = {W \mathord{\left/  {\vphantom {W {\left( {N - 1} \right),}}} \right. \kern-\nulldelimiterspace} {\left( {N - 1} \right)}}$, $N > 1$ denotes the physical spacing between adjacent ports in units of the wavelength.
The reasons of parameter selection is provided as follows.
For the probability interval $\left[ {{{10}^{ - 6}}, {{10}^{ - 2}}} \right]$, the choice of threshold $u$ specifies the upper probability bound for the GPD fitting, i.e.,  ${{\hat F}_{{\rm{upper}}}} = {F_{{\rm{GEV}}}}\left( u \right) = {10^{ - 2}}$. The lower probability bound ${{\hat F}_{{\rm{lower}}}} = {10^{ - 6}}$ is set in accordance with the reliability requirements of ultra-reliable communications, where OPs on the order of $10^{-6}$ are typically of interest.
For the system parameters ranges, the detailed reason has been explained in \cite{Xu2025PerformanceAO} from both the physical size constraints and the performance plateau phenomenon of FAS.
Under the above parameter settings, the fitting results of GPD parameters (${{\xi _u}}$, ${{\sigma _u}}$) using MMSE criterion are presented in \eqref{4A} at the bottom of this page.
It is worth noting that the GEV parameters (${\xi}$, ${\sigma}$, ${\mu}$) have been obtained using ML criterion, and the relationships between (${\xi}$, ${\sigma}$, ${\mu}$) and $\left( {W,N} \right)$ have been derived in (29)--(31) of \cite{Xu2025PerformanceAO}, as reproduced in \eqref{5A}.
\begin{align}
{\xi _u} \approx  &-0.2323 + 6.225\times {10^{ - 2}}W +1.136\times {10^{ - 3}}N\notag\\
& -1.067\times {10^{ - 2}}{W^2} - 7.649\times {10^{ - 4}}WN\notag\\
 &-3.942 \times{10^{ - 8}}{N^2} + 2.36\times {10^{ -4}}{W^2}N\notag\\
& - 1.163\times {10^{ - 5}}W{N^2} +3.076 \times{10^{ -7}}{N^3},\notag\\
{\sigma _u} \approx &0.07778+ 1.17\times {10^{ - 2}}W + 5.904\times {10^{ - 4}}N \notag\\
&-1.629\times {10^{ - 3}}{W^2} - 4.019\times {10^{ - 4}}WN \notag\\
&+6.439 \times{10^{ - 6}}{N^2} + 4.79\times {10^{ -5}}{W^2}N \notag\\
&+  3.224 \times {10^{ - 7}}W{N^2} - 4.909 \times{10^{ -8}}{N^3}.\label{4A}
\end{align}
\vspace{-1.5em}
\begin{align}
{\xi}  \approx  &-0.1235 + 1.014\times {10^{ - 3}}W -  8.942\times {10^{ - 6}}N\notag\\
 &+7.796\times {10^{ - 4}}{W^2} - 8.619\times {10^{ - 5}}WN\notag\\
  &+1.867 \times{10^{ - 6}}{N^2} + 1.867\times {10^{ -6}}{W^2}N\notag\\
  & +  2.332 \times {10^{ - 6}}W{N^2} - 6.288 \times{10^{ -8}}{N^3},\notag\\
{\sigma}  \approx &0.4039 - 3.814\times {10^{ - 2}}W + 8.851\times {10^{ - 4}}N\notag\\
 &+3.338\times {10^{ - 3}}{W^2} + 3.779\times {10^{ - 4}}WN\notag\\
  &- 2.798 \times {10^{ -5}}{N^2}- 5.65 \times {10^{ -5}}{W^2}N\notag\\
   &+ 1.552 \times {10^{ - 6}}W{N^2} + 1.004 \times{10^{ - 7}}{N^3},\notag\\
{\mu}  \approx &0.9346 + 2.511\times {10^{ - 1}}W + 9.196\times {10^{ - 3}}N\notag\\
 &- 3.177\times {10^{ - 2}}{W^2}-6.431\times {10^{ - 4}}WN\notag\\
 & - 1.44 \times {10^{ - 4}}{N^2} + 4.325\times {10^{ - 4}}{W^2}N\notag\\
  & - 2.548\times {10^{ -5}}W{N^2} + 1.404 \times {10^{ - 6}}{N^3}.\label{5A}
\end{align}

The generalization of these parameters inevitably introduces minor approximation errors, while they have a negligible impact on the OP within the fitted parameter range. Moreover, in practical engineering applications, FA are typically manufactured with only a limited set of standardized $W$ and $N$ values. For these discrete configurations, a lookup table of precise parameters for target distribution can be provided, enabling a more accurate OP evaluation in practical systems. The detailed parameter table can be obtained by following the proposed fitting procedure and is omitted here for brevity.

\emph{OP Evaluation:} With $\left( {\xi} ,{\sigma},{\mu } \right)$  and (${{\xi_u}}$, ${{\sigma _u}}$), the approximate distribution of $\left| {{h_{{\rm{FAS}}}}} \right|$ can be obtained by substituting these parameters into \eqref{3A}. Based on the fitted channel distribution, the OP can then be evaluated by setting $x=\hat{\gamma}$. It is worth noting that the fitted parameter expressions in \eqref{4A} and \eqref{5A} are derived under the assumption of unit-average-power small-scale fading, i.e., ${\sigma ^2} = 1$. Nevertheless, the fitted FAS channel distribution remains applicable to arbitrary small-scale fading power and large-scale fading conditions due to the scaling property $\Pr \left( {\sqrt {{d^{ - \alpha }}{\sigma ^2}} \left| {{h_{{\rm{FAS}}}}} \right| \le {\hat{\gamma}}} \right) = \Pr \left( {\left| {{h_{{\rm{FAS}}}}} \right| \le \frac{{\hat{\gamma}}}{{\sqrt {{d^{ - \alpha }}{\sigma^2}} }}} \right)$.

It should be emphasized that the quality of the polynomial approximation cannot be fully assessed solely through parameter-level fitting metrics. Instead, its effectiveness should be evaluated from a system-performance perspective. In particular, since the ultimate objective of the proposed EVD-based fitting framework is the accurate OP characterization, even small parameter deviations may result in non-negligible OP errors, especially in the ultra-low-probability region. Therefore, the effectiveness of the proposed approximation will be further evaluated through OP analysis in Section \ref{sec:Results}.


\begin{figure}
\centering
\includegraphics[width=0.85\columnwidth]{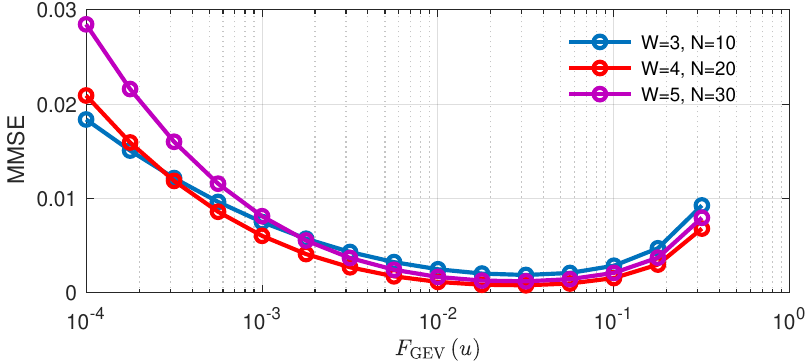}\\
\caption{MMSE of OP versus ${F_{{\rm{GEV}}}}\left( u \right)$ for threshold selection.}\label{F1}
\end{figure}

\begin{figure}
\centering
\includegraphics[width=0.85\columnwidth]{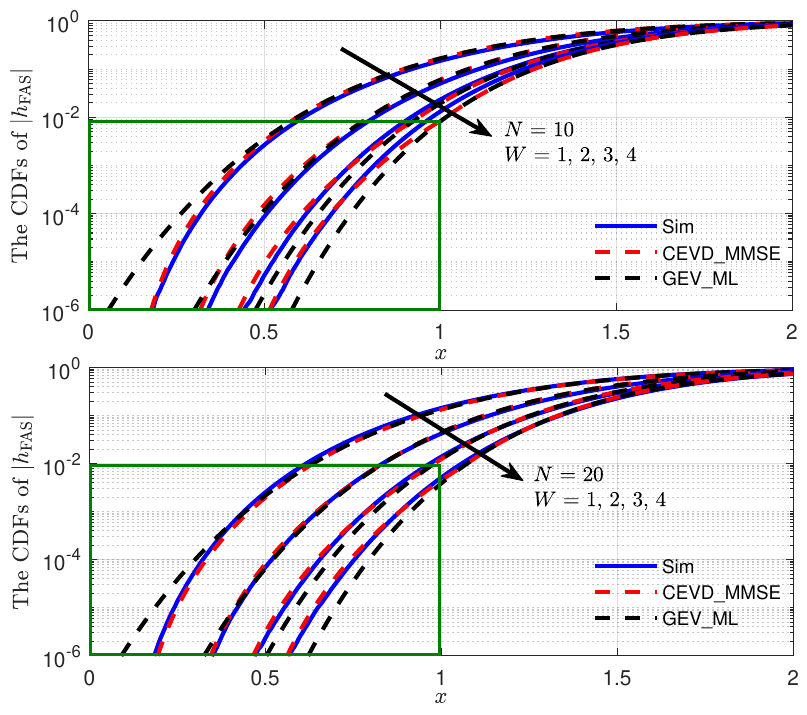}\\
\caption{The CDFs of $|h_{\rm FAS}|$.}\label{F2}
\vspace{-1em}
\end{figure}

\section{Numerical Results}\label{sec:Results}
In this section, Monte-Carlo simulation results serve as the benchmark, against which we compare the proposed OP-oriented fitting strategy and the ML-based GEV fitting strategy in \cite{Xu2025PerformanceAO}.
In the Monte-Carlo simulations, $10^8$ independent channel realizations of $|h_{\rm FAS}|$ are generated to estimate the empirical CDF. At the OP level of $10^{-6}$, the relative standard deviation of the Monte-Carlo estimate is approximately $10\%$.
In all figures, Monte-Carlo simulations are labeled as ``Sim", the ML-based GEV fitting strategy as ``GEV\_ML", and the MMSE-based CEVD fitting strategy as ``CEVD\_MMSE". The parameters used in the target distributions are obtained based on the generalized expressions in \eqref{4A} and \eqref{5A}.

Fig. \ref{F1} shows the MMSE, calculated according to \eqref{7A}, versus ${F_{{\rm{GEV}}}}\left( u \right)$ for threshold selection. The MMSE first decreases and then increases as ${F_{{\rm{GEV}}}}\left( u \right)$ grows, which reflects the impact of $u$. Moreover, the MMSE remains relatively low and stable around ${F_{{\rm{GEV}}}}\left( u\right)={10^{ - 2}}$, demonstrating the reasonableness of ${F_{{\rm{GEV}}}}\left( u\right)\!=\!{10^{ - 2}}$.

 %

Fig. \ref{F2} shows the CDFs of $\left| {{h_{{\rm{FAS}}}}} \right|$ for different values of $W$ under $N=10$ and $N=20$ using various fitting strategies.
The dark-green rectangular regions highlight where
the proposed CEVD\_MMSE fitting strategy achieves more accurate fitting of the FAS channel distribution in the left-tail region than the GEV\_ML strategy.
We observe that the GEV\_ML exhibits observable deviations in the ultra-low-probability region, whereas the proposed strategy effectively mitigates these discrepancies. Although parameter generalization in \eqref{4A} and \eqref{5A} introduces minor approximation errors, their impact on the fitting accuracy can be negligible.

\begin{figure}
\centering
\includegraphics[width=0.85\columnwidth]{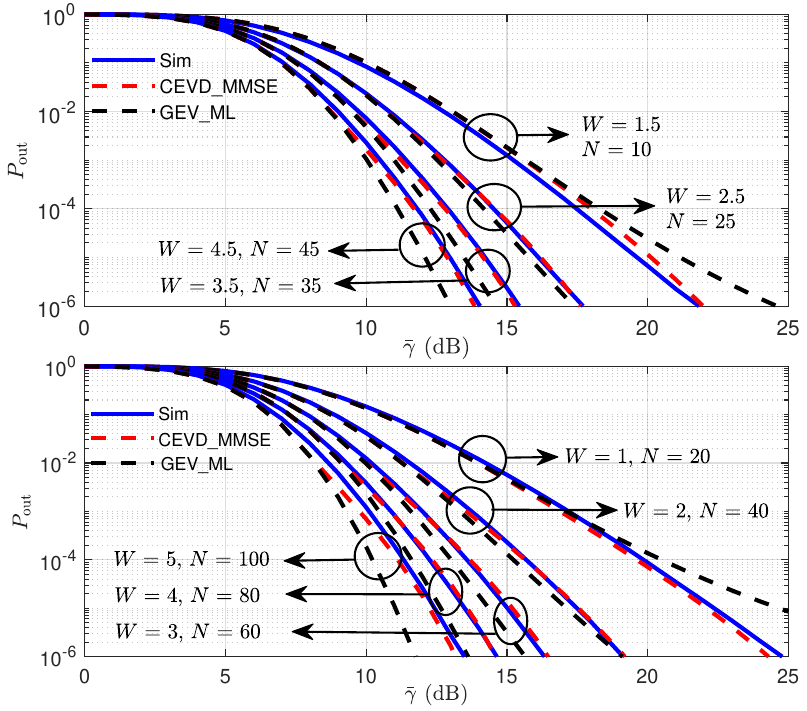}\\
\caption{The OPs of FAS versus $\bar \gamma$.}\label{F3}
\vspace{-1em}
\end{figure}

\begin{figure}
\centering
\includegraphics[width=0.85\columnwidth]{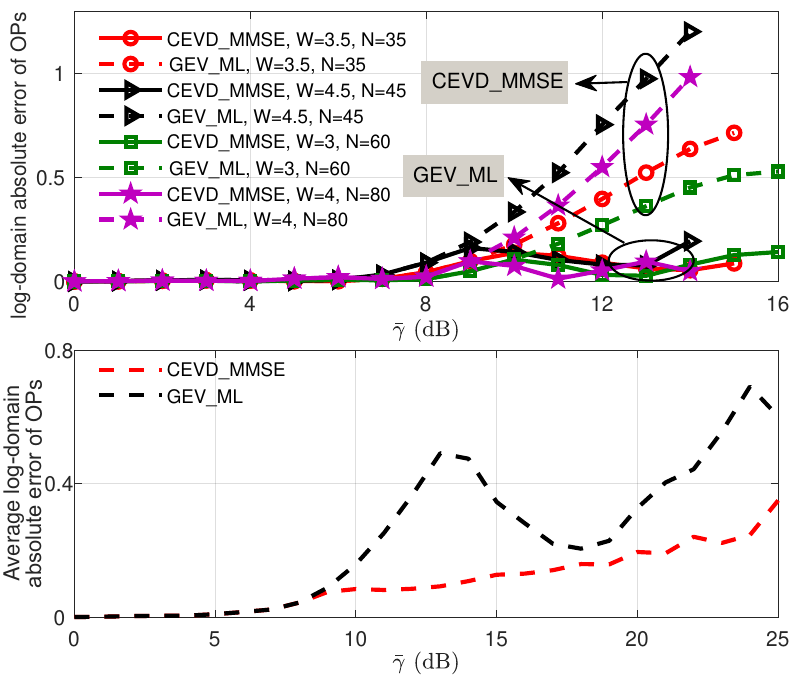}\\
\caption{The log-domain absolute errors of OPs versus $\bar \gamma$.}\label{F4}
\end{figure}

Fig. \ref{F3} illustrates the approximate OPs versus transmit SNR $\bar{\gamma}$ with $\gamma_{\rm{th}}=10$ dB under moderate-density port deployment (top panel) and dense port deployment (bottom panel). We observe that the proposed CEVD\_MMSE strategy effectively eliminates the inaccuracies exhibited by GEV\_ML in the high-SNR region for both deployment densities, thereby verifying the effectiveness of the proposed strategy.

\begin{figure}[!t]
\centering
\includegraphics[width=0.85\columnwidth]{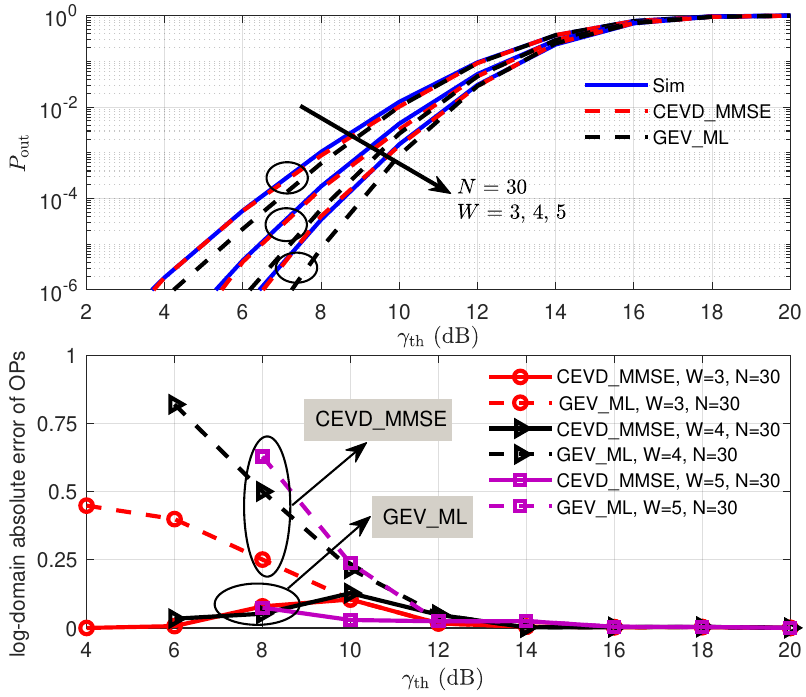}\\
\caption{The OPs and log-domain absolute errors versus $ \gamma_{\rm{th}}$.}\label{F5}
\vspace{-1em}
\end{figure}
To further quantify the fitting gap between the proposed strategy and GEV\_ML with respect to Monte-Carlo simulation results, Fig. \ref{F4} presents the log-domain absolute error $\left| {{{\log }_{10}}\left( {P_{{\rm{out}}}^{{\rm{sim}}}} \right) - {{\log }_{10}}\left( {P_{{\rm{out}}}^{{\rm{fit}}}} \right)} \right|$ (top panel) under specific system settings, as well as the average log-domain absolute error (bottom panel) under $W \in \left[ {0.5,5} \right]$ and $\rho  \in \left[ {0.05,0.5} \right]$, both versus transmit SNR. It is evident that the errors associated with the proposed strategy are smaller than those of GEV\_ML.

Fig. \ref{F5} shows the approximate OPs and log-domain absolute error versus the decoding SNR threshold $\gamma_{\rm{th}}$ with $\bar{\gamma}=10$ dB. The results demonstrate that CEVD\_MMSE achieves more accurate OP evaluation in the low decoding threshold region.

\section{Conclusions}\label{sec:conclude}
This paper proposed an OP-oriented FAS channel fitting strategy, where a CEVD and a MMSE criterion, both tailored to emphasize the left-tail region, are developed. Simulation results show that the proposed strategy significantly improves FAS channel fitting accuracy in the left-tail region, thereby enhancing OP evaluation accuracy in the low-outage region. This highlights the importance of performance-oriented channel modeling, where the fitting strategy should be explicitly designed according to the target performance metric.
\ifCLASSOPTIONcaptionsoff
  \newpage
\fi
\bibliographystyle{IEEEtran}
\bibliography{refa}

\end{document}